# Weak Measurements and Counterfactual Computation

**Onur Hosten and Paul Kwiat**

*Department of Physics, University of Illinois at Urbana Champaign, Urbana, Illinois 61801-3080*

Vaidman, in his article [1] first gives a very clear and accurate description of the existing procedures for counterfactual computation (CFC), where one obtains information from a quantum computer, without actually running the computer [2]. Following that, he adopts the method of "quantum weak measurements in pre- and post-selected ensembles" to ascertain whether or not the chained-Zeno CFC scheme in [2] is counterfactual; which has been the topic of a debate on the definition of counterfactuality [3, 4]. We disagree with his conclusion, which brings up some interesting aspects of quantum weak measurements [5] and some concerns about the way they are interpreted, particularly in [1].

The summarizing question of the debate is as follows: Has a single photon entering the concatenated interferometers in Fig. 1 passed through the computer (comp) (or equivalently through path B) or not, in the outcome in which it gets detected at path D? Our argument concerning the answer goes as follows: Given the fact that the path lengths of the inner interferometer, formed by the two beamsplitters (BS), are adjusted such that the amplitudes traveling through paths B and C give complete destructive interference at the interferometer output leading to path F, no photon actually passes through path F. This can be verified by placing a detector at F. Then, the only way a photon can reach path D is if it is coming from path A. Therefore, a photon does *not* pass through path B, or the computer, before arriving at path D.

How might we confirm this? Any *strong* measurement at point B to determine whether or not the photon was there before it is detected at path D will collapse the coherent evolution of the system wavefunction, eliminating the destructive interference condition on path F. Therefore a strong measurement will change the fact which path the photon takes [6]. However, one can imagine making very weak measurements, which give only very little which-path information, still allowing interference to take place. This type of measurement in between an initial and a final condition, i.e., between an initial and a final strong measurement, is known as "weak measurements in pre- and post-selected ensembles", and its outcome is called the "weak-value" of the measured observable [5]. Since [1] relies on the outcome of a weak measurement analysis, we explain such an analysis here, so that we can then discuss its limitations. In our case, the pre-selection is that the photon starts at the top left in Fig. 1, and the post-selection is that the photon ends up at path D. Since the measurements are weak, a measurement on one photon does not give much information, and one needs to repeat the experiment many times, and obtain an ensemble average to resolve the weak-value. In the limit that the measurement strength goes to zero (and number of repetitions goes to infinity), the weak-values of two non-commuting observables can be measured simultaneously. Particularly, a weak measurement to see if the photon is on path B does not affect the outcome of the result of the weak measurement performed on path F, in contrast to the case of strong measurements.

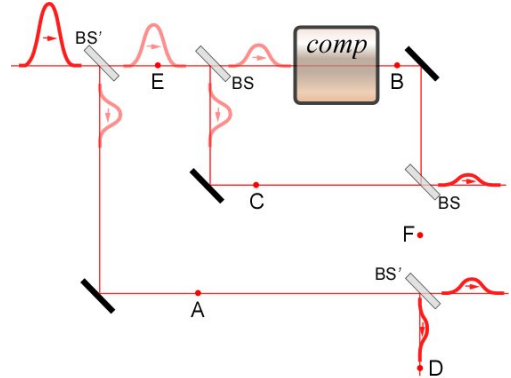

**Figure 1** The pulse representing a single-photon probability distribution traveling through the concatenated interferometers.

The simplest optical implementation of such a measurement would be to use the transverse spatial distribution of the photon wave packet as the 'meter', and entangle the observable to be measured to this 'meter'. In our case, the observable is the path of the photon, i.e., a projection onto a path (weak). Entangling the meter and the observable could be realized, e.g., by placing a tilted parallel glass slab into the path to be measured in Fig. 1, so that the transverse spatial distribution of the photon will shift slightly towards one side, only if the photon takes that particular path [7]. The shift at the end of the experiment – the weak-value – is then to be read out on path D. The amount of shift, of course, has to be much smaller than the width of the beam itself for the measurement to be a weak one.

The weak-values of projections can be called weak-probabilities, which form a quasi-probability distribution when one considers a complete set of projectors, which can have negative elements (similar to the well-known Wigner function). In our case, measuring a negative value means that we displaced the beam one way on a certain path, but the beam ended up shifting the opposite way on path D (due to an interference effect of the amplitudes between various paths).

The weak-value $A_w$ of an observable **A** in the first-order approximation is given by $A_w=\langle\psi_2|\mathbf{A}|\psi_1\rangle/\langle\psi_2|\psi_1\rangle$, and can be outside of the range of its eigenvalues [5]. Here $|\psi_1\rangle$ is the pre-selected state and $|\psi_2\rangle$ is the post-selected state. The only difference from a regular expectation value is the presence of $\langle\psi_2|$ in place of $\langle\psi_1|$. Physically, $A_w$ corresponds to the amount of shift in the central position of

the beam up to a constant multiplier defined by the interaction between the system and the meter. Since we take the limit of an infinitely weak measurement, higher-order terms contributing to the shift are negligibly small in comparison to the first order. This is why the weak-values of two non-commuting observables can be measured simultaneously.

Here, the observables are the projection operators $|i\rangle\langle i|$, which have two possible eigenvalues, 0 and 1 (i can range from path A to path F in Fig. 1). The weak-probabilities of finding the photon in various paths, and also the assertions that these weak-probabilities make about a single photon traversing the circuit, given the pre- and post-selection succeeds, are as follows:

$E_w=0$ : photon never enters the inner interferometer,
$F_w=0$ : photon never leaves the inner interferometer,
$A_w=1$ : weak probability of the photon being on path A is 1,
$B_w=1$ : weak probability of the photon being on path B is 1,
$C_w=-1$ : weak prob. of the photon being on path C is $-1$,
$B_w+C_w=0$ : The photon is not in the inner interferometer (consistent with $E_w=0$ and $F_w=0$),
$A_w+B_w+C_w = A_w+E_w = A_w+F_w=1$: total probability adds up to 1 at any given instant, as it should.

If interpreted literally, we will arrive at the conclusions made by Vaidman in [1]:
"weak measurement requires a pre- and post-selected ensemble, but according to Hosten *et al.*, for all members of the ensemble the photon is not in B, so we should not see any effect in the measurement in B. The experiment, however, will show a different result: the outcome of the weak measurement of the projection inside the "computer" is 1……The photon did not enter the interferometer, the photon never left the interferometer, but it was there! This is a new paradoxical feature of a pre- and post-selected quantum particle."

Therefore, Vaidman concludes that, given the properties of weak measurements, since the outcome of $B_w$ is not 0, the photon actually passes through the computer. We disagree, and say that the photon is only 'weakly' present on path B, and this is only because of the weak measurements performed on path B, i.e., no matter how weak, the weak measurements disturb the system, and disturb the counterfactuality.

Even in the limit where the measurement strength goes to zero, what gives rise to a finite $B_w$ is the amplitude leaking through path F caused by the disturbance of the 'weak' measurement on path B. That is, the infinitesimal displacement of the beam on path B due to the insertion of an infinitesimally tilted glass slab changes the mode-matching on the second BS of the inner interferometer, and there is no longer perfect destructive interference on path F. Consequently, there is some leaking amplitude on path F. This leaking amplitude is what causes the shift on path D (by interfering with the amplitude coming from path A), where the central position of the beam is read out. Moreover, resolving the central position requires that we repeat the experiment many times, such that >>1 photon leaks through path F on the average; these photons in turn come from the computer, *because* of the weak measurement.

Therefore, in the presence of weak measurements (no matter how weak) it is not surprising to see some presence of photons on path B in the outcome which the photons are finally detected on path D. Moreover, in the presence of weak measurements, there is an unavoidable actual photon flux (on the average >>1 photon during the entire experiment) into and out of the inner interferometer then to path D, in contrast to Vaidman's interpretation.

One might ask, why do weak measurements performed on path F give the result $F_w=0$, regardless of whether or not we make weak measurements on path B? Our calculations show that, if we take into account the *higher-order corrections*, $F_w$ is *non*zero when we make weak measurements on path B; however, in the limit that the measurement strength goes to zero, these corrections go to zero as well (unlike the first-order terms), at the expense of increasing the number of repetitions necessary to resolve the weak-value. Consequently, weak measurements do not 'see' the photons leaking through path F.

We conclude that, no matter how weak, weak measurements disturb the counterfactuality. Only in the absence of any kind of known quantum measurements is the protocol in Fig. 1 (in [1] and [2] also) counterfactual. We would also like to acknowledge that this work raises some questions about the interpretation of the three-box problem via weak measurements (depending on the setup implementing it), which can have similarities with the current discussion as shown in [1].